\documentclass[letterpaper]{article}
\usepackage[preprint]{aaai2027arxiv}
\usepackage[hyphens]{url}
\usepackage{graphicx}
\usepackage{natbib}
\usepackage{caption}
\usepackage{booktabs}
\usepackage{multirow}
\usepackage{array}
\usepackage{amsmath}
\usepackage{xcolor}
\usepackage{colortbl}
\definecolor{gFieldD}{HTML}{2F7D32}\colorlet{gField}{gFieldD!14}
\definecolor{gStapleD}{HTML}{B7791F}\colorlet{gStaple}{gStapleD!16}
\definecolor{gFruitD}{HTML}{B03A5B}\colorlet{gFruit}{gFruitD!12}
\definecolor{gVegD}{HTML}{1F6E9C}\colorlet{gVeg}{gVegD!12}
\definecolor{gMeanD}{HTML}{4A4A4A}\colorlet{gMean}{gMeanD!10}
\definecolor{oursrow}{RGB}{236,242,250}
\usepackage{pifont}
\definecolor{bestgreen}{RGB}{200,240,205}
\newcommand{\cm}{\ding{51}}
\newcommand{\dm}{--}
\newcommand{\mcite}[2]{#1 {\scriptsize\citep{#2}}}
\title{MACE: Memory-Augmented Crop-Disease Evaluation\\of Frozen Vision-Language Models}

\author{
Tirtho Roy\textsuperscript{\rm 1}\equalcontrib,
Shreyan Ganguly\textsuperscript{\rm 1}\equalcontrib,
Muhammad Arbab Arshad\textsuperscript{\rm 1},
Nikhil Bhattacharya\textsuperscript{\rm 1},
\mbox{Soumik Sarkar\textsuperscript{\rm 1}}
}
\affiliations{
\textsuperscript{\rm 1}Iowa State University, Ames, IA, USA\\
tirtho@iastate.edu, soumiks@iastate.edu
}

\usepackage[colorlinks=true,linkcolor=blue!55!black,citecolor=blue!55!black,urlcolor=blue!55!black,pdftitle={MACE: Memory-Augmented Crop-Disease Evaluation of Frozen Vision-Language Models},pdfauthor={Tirtho Roy, Shreyan Ganguly, Muhammad Arbab Arshad, Nikhil Bhattacharya, Soumik Sarkar}]{hyperref}
\begin{document}

\maketitle

\begin{abstract}
In plant phenotyping, crop disease diagnosis has traditionally relied on expert assessment of visible symptoms and on models tailored to specific crops or diseases, and both are difficult to scale across field conditions. Recent methods provide frozen vision-language models with source-grounded knowledge bases that summarize the typical symptoms of each disease but describe little of how symptoms vary across individual cases. We introduce \textbf{MACE} (Memory-Augmented Crop-Disease Evaluation), a training-free method that supplements the knowledge base with a memory of symptom descriptions paired with known disease labels. MACE matches a description of the query image against this memory and retrieves only text descriptions and disease labels, so the query is the only image that the diagnosis model sees. Across ten crops, MACE with Claude Sonnet 5.5 improved accuracy over the knowledge-base baseline from 42.8\% to 62.2\% in distribution and from 50.7\% to 57.0\% on held-out image collections. Retrieved disease labels accounted for most of the gain, and adding the corresponding text descriptions improved in-distribution accuracy by about six percentage points.
\end{abstract}

\begin{links}
  \link{Website and code}{https://mace-bench.github.io/}
  \link{Dataset}{https://hf.co/datasets/tirtho149/MACE-Bench}
\end{links}

\section{Introduction}

Food security remains a global challenge, with 2.1 billion people experiencing moderate or severe food insecurity worldwide \citep{fao2026sofi}. Timely crop disease diagnosis can help protect agricultural productivity, particularly where access to expert scouting is limited. However, task-specific neural networks require labeled disease images and often generalize poorly beyond the collections used for training \citep{xu2023limited,wu2023field,xu2024datasets}.

Frontier vision-language models (VLMs) offer a flexible alternative because broad pretraining and in-context learning allow them to address new tasks with few labeled examples. In agriculture, instruction-tuned models \citep{liu2024multimodal,awais2025agrogpt}, broad benchmark evaluations \citep{shinoda2025agrobench,gauba2025agmmu}, and adaptation through labeled examples \citep{arshad2025leveraging} have been studied (Table~\ref{tab:prior}). AgEval \citep{arshad2025leveraging} showed that VLMs can adapt to plant stress phenotyping without task-specific training, while AgroBench \citep{shinoda2025agrobench} showed that fine-grained agricultural recognition remains challenging across expert-annotated tasks. For crop disease diagnosis, SAGE provides a frozen VLM with candidate diseases, source-grounded symptom knowledge, and labeled reference images \citep{arshad2026sage}.

While reference images provide useful case-specific evidence, processing them during diagnosis increases inference cost, and reference pools drawn from test collections can contain near-copies \citep{lee2022deduplicating,abbas2023semdedup}. In such cases, accuracy may reflect label lookup rather than diagnosis, as documented for data leakage and dataset-specific shortcuts \citep{kapoor2023leakage,xu2024datasets}. A source-grounded symptom knowledge base (KB) avoids image matching, but it describes each disease at the class level and does not describe how symptoms vary across individual labeled cases. Retrieval-augmented models add retrieved cases or documents to the input at inference time, and recent work applies this idea to plant disease diagnosis and to disease classification with multimodal large language models \citep{chan2026smart,zhan2026raicl}. A caption memory uses the same principle with text alone: it describes each reference image once and reuses the instance-level symptoms and the disease label without processing the reference image again or comparing it with the query image.

\begin{figure*}[t]
\centering
\includegraphics[width=\textwidth]{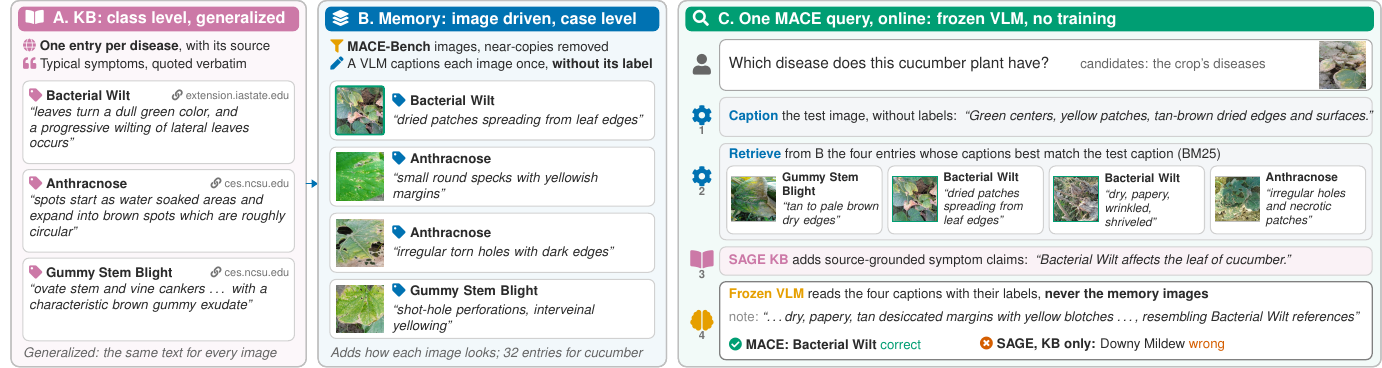}
\caption{\textbf{MACE on a recorded query.} (A) The KB holds one generalized entry per disease. (B) The memory holds image-driven captions with labels. (C) MACE retrieves four memory entries and adds KB claims, and a frozen VLM answers.}
\label{fig:overview}
\end{figure*}

\begin{table}[t]
\centering\scriptsize
\caption{\textbf{Prior work} by evidence and evaluation setting. \cm: reported; \dm: not reported.}
\label{tab:prior}
\resizebox{\columnwidth}{!}{\begin{tabular}{@{}lcccccc@{}}
\toprule
\textbf{Method} & \textbf{Labeled} & \textbf{Source} & \textbf{Query} & \textbf{KB} & \textbf{Cited} & \textbf{Text} \\
 & \textbf{refs} & \textbf{shift} & \textbf{retrieval} &  & \textbf{evidence} & \textbf{memory} \\
\midrule
\mcite{CDDM}{liu2024multimodal} & \dm & \dm & \dm & \dm & \dm & \dm \\
\mcite{AgroGPT}{awais2025agrogpt} & \dm & \cm & \dm & \dm & \dm & \dm \\
\midrule
\mcite{AgroBench}{shinoda2025agrobench} & \dm & \dm & \dm & \dm & \dm & \dm \\
\mcite{AgMMU}{gauba2025agmmu} & \dm & \dm & \dm & \dm & \dm & \dm \\
\midrule
\mcite{Tip-Adapter}{zhang2022tip} & \cm & \cm & \cm & \dm & \dm & \dm \\
\mcite{AgEval}{arshad2025leveraging} & \cm & \dm & \dm & \dm & \dm & \dm \\
\mcite{SAGE}{arshad2026sage} & \cm & \cm & \cm & \cm & \cm & \dm \\
\midrule
\textbf{MACE (ours)} & \cm & \cm & \cm & \cm & \cm & \cm \\
\bottomrule
\end{tabular}}
\end{table}

We introduce \textbf{MACE} (Memory-Augmented Crop-Disease Evaluation), a training-free method that adds a \emph{caption memory} of labeled reference images to the symptom KB of SAGE (Figure~\ref{fig:overview}). A VLM describes each reference image once, without its label, in structured visual fields, and MACE stores the description together with the disease label. At test time, MACE describes the query image in the same way, retrieves the most similar memory entries, and provides their descriptions and labels to a frozen diagnosis VLM, which never sees the memory images.

The main contributions of this work are as follows:
\begin{enumerate}
\item We introduce MACE, a training-free method that describes each labeled reference image with a structured, label-free caption and retrieves these captions with their disease labels for a frozen diagnosis VLM.
\item We establish an evaluation across ten crops, two frozen diagnosis models, and held-out image collections, with near-duplicate removal and comparisons against source-grounded diagnosis and visual nearest-neighbor lookup.
\item We analyze how caption content, retrieved labels, memory construction, and repeated disease evidence contribute to diagnosis, and we find that retrieved labels account for most of the gain.
\item We introduce a memory retriever whose weights a large language model (LLM) writes without gradient training, and we compare it with fitted, lexical, and visual selection rules on both coverage and diagnosis accuracy.
\end{enumerate}

\section{Method}

\paragraph{Data.} We use the crop-disease images, labels, and KB entries compiled by \citet{arshad2026sage}. We remove exact and perceptual duplicates, images with visible disease names, and within-disease near-copies with an OpenCLIP cosine similarity of at least 0.90 \citep{cherti2023reproducible,abbas2023semdedup}. For each crop, we hold out one complete source collection and split the remaining images into five folds. For each of three memory selections, we choose $K$ images per disease from the training folds with a facility-location objective, and the selections are nested across memory sizes.

\begin{figure*}[t]
\centering
\includegraphics[width=\textwidth]{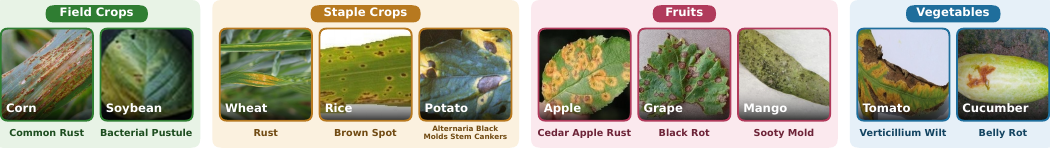}
\caption{\textbf{The ten evaluated crops} by crop group, each shown with one test image and its disease label.}
\label{fig:cropset}
\end{figure*}

\paragraph{Caption memory.} For each reference image, we use Claude Sonnet to generate a structured, label-free caption from visible symptoms alone. We store this caption with the image's known disease label as one caption-memory entry (Figure~\ref{fig:overview}B). At test time, we use the same prompt to caption the query image, while the diagnosis model receives only memory captions and labels and never sees the reference images.

\paragraph{Compared methods.} We pair the caption memory with two frozen frontier models from the Claude family, Sonnet and Haiku (\texttt{claude-sonnet-5-5}, \texttt{claude-haiku-5-5}), as diagnosis models. Sonnet is evaluated under every condition, while Haiku is evaluated under SAGE, MACE retrieval, and labels only. Because Haiku reads the same captions written by Sonnet, the Haiku runs differ from the Sonnet runs only in the diagnosis model. We send 512-pixel images to both models through the Claude Code command-line interface with default decoding, and we disable prompt caching and external tools. \emph{SAGE}, the KB-only baseline, receives the test image, the candidate labels, and KB claims (up to 3,000 tokens) ranked against a label-free caption of the test image, and \emph{MACE retrieval} adds the retrieved memory entries described above. The KB, candidate labels, prompts, and output schema stay fixed, so the content controls differ from MACE only in the memory content: \emph{labels only} omits captions, \emph{captions only} omits labels, \emph{shuffled labels} permutes labels, and \emph{random references} presents random memory entries. Label lookup does not use a VLM and predicts either by a majority or score-weighted vote over the retrieved labels or by the label of the nearest memory image (1-NN) under OpenCLIP or DINOv2 \citep{oquab2024dinov2} similarity. Table~\ref{tab:hero} lists these methods with their accuracy per crop and their equal-crop mean on the full evaluation. Unless stated otherwise, intervals are paired, crop-stratified 95\% bootstrap intervals, $p$-values are Holm-corrected within each family of contrasts, and paired answer changes are tested with exact McNemar tests.

\paragraph{Ablations.} The labels-only ablation removes the caption text from MACE, and we evaluate it on the full query set with both diagnosis models. In addition, a \emph{DINOv2-only} ablation retrieves the four highest-ranked DINOv2 entries from the same memory instead of using BM25, the default caption retrieval described below.

\paragraph{Selection rules.} A selection rule determines which four memory entries the diagnosis model sees for a query. The rule scores every entry in the query's memory against the query and keeps the four highest. By default, MACE uses BM25 \citep{robertson2009probabilistic}, which ranks memory captions by the words they share with the query caption and weights words that are rare in the memory more heavily. Our BM25 implementation reproduces the recorded selection for every query, so the recorded runs serve as its baseline.

\paragraph{Alternative rules.} Each alternative changes only the scoring step, while one memory selection, the diagnosis model, the prompt, the KB, and the four-entry format stay fixed. \emph{Text-embedding cosine} compares frozen sentence embeddings of the two captions, and \emph{rank fusion} merges the BM25 and cosine rankings. \emph{BM25 + DINOv2} adds the image similarity between the query and each memory image, rescaled within the memory, to a retuned BM25 score, and the diagnosis model still receives only captions and labels. The fitted scorers combine five caption features (shared rare words, word overlap, character overlap, embedding similarity, and length ratio) through a linear model or a multilayer perceptron (MLP) with 29 parameters trained on labeled pairs. These supervised scorers are comparison baselines and are not part of the training-free MACE method. In the \emph{LLM-written MLP}, Sonnet writes the weights of the same MLP without gradient training, given a description of the features alone or together with 16 memory captions, 16 labeled example triplets, or both. The architecture, features, and prompt were fixed during development.

\paragraph{Tuning and coverage.} We selected every tuned parameter and generated weight on 2,215 training-fold images that are disjoint from all evaluation queries and from the memory against which they are scored, and we fixed these values before testing. We measure coverage with hit@4, the share of queries for which at least one of the four references is labeled with the true disease.

\paragraph{Crops and queries.} We selected ten crops from metadata before examining comparative accuracy. The crops span field crops, staple crops, fruits, and vegetables (Figure~\ref{fig:cropset}). Soybean, corn, and tomato were fixed in advance, and the remaining crops are the eligible ones with the most held-out diseases. A crop is eligible if it has a held-out collection, at least 60 held-out images, four in-distribution (IID) diseases, and two IID collections. The held-out collection is Bugwood for seven crops and LeafNet, SBRD, and CDDM for mango, rice, and tomato, respectively. Each crop contributes up to 100 IID and 60 held-out queries drawn round-robin across diseases, and a query never appears in its own memory. Table~\ref{tab:hero} reports the first memory selection; with the other two selections, Sonnet MACE accuracy changes by less than three points. In total, the full evaluation contains 1,600 queries, the nested ablation set contains 400, and a targeted selection check uses 200 ablation queries from five crops.

\begin{table*}[t]
\centering
\caption{\textbf{Accuracy (\%) per crop} on in-distribution (IID) and held-out (HO) queries ($K{=}4$). $\Delta$: points from Sonnet SAGE (bootstrap, Holm; $^{*}p{<}0.05$, $^{**}p{<}0.01$, $^{***}p{<}0.001$). $^{\dagger}$: Bugwood held-out. Green bold: best; underlined: second.}
\label{tab:hero}
\resizebox{\textwidth}{!}{\footnotesize
\begin{tabular}{cll|cc|ccc|ccc|cc|c|r}
\toprule
 & & & \multicolumn{2}{|c}{\cellcolor{gField}\textcolor{gFieldD}{\textbf{Field}}} & \multicolumn{3}{|c}{\cellcolor{gStaple}\textcolor{gStapleD}{\textbf{Staple}}} & \multicolumn{3}{|c}{\cellcolor{gFruit}\textcolor{gFruitD}{\textbf{Fruits}}} & \multicolumn{2}{|c}{\cellcolor{gVeg}\textcolor{gVegD}{\textbf{Vegetables}}} & \multicolumn{1}{|c}{\cellcolor{gMean}\textcolor{gMeanD}{\textbf{Mean}}} & \multicolumn{1}{|c}{\cellcolor{gMean}\textcolor{gMeanD}{\textbf{$\Delta$ vs baseline}}} \\
 & Method & VLM & \cellcolor{gField}Corn$^{\dagger}$ & \cellcolor{gField}Soybean$^{\dagger}$ & \cellcolor{gStaple}Wheat$^{\dagger}$ & \cellcolor{gStaple}Rice & \cellcolor{gStaple}Potato$^{\dagger}$ & \cellcolor{gFruit}Apple$^{\dagger}$ & \cellcolor{gFruit}Grape$^{\dagger}$ & \cellcolor{gFruit}Mango & \cellcolor{gVeg}Tomato & \cellcolor{gVeg}Cucumber$^{\dagger}$ & \cellcolor{gMean}All crops & \cellcolor{gMean}SAGE, Sonnet \\
\midrule
\multicolumn{15}{l}{\cellcolor{gMean}\textbf{In-distribution (IID) queries}} \\
 & Label majority vote & -- & 32.0 & 23.0 & 63.0 & 54.0 & 76.0 & 60.0 & 56.0 & 66.0 & 25.0 & \underline{71.0} & 52.6 & +9.8$^{***}$ \\
 & Score-weighted vote & -- & 35.0 & 37.0 & 65.0 & 52.0 & 74.0 & 60.0 & 59.0 & \underline{72.0} & 29.0 & \cellcolor{bestgreen}\textbf{74.0} & 55.7 & +12.9$^{***}$ \\
 & OpenCLIP 1-NN & -- & \underline{38.0} & 34.0 & 60.0 & \underline{56.0} & 73.0 & 59.0 & \cellcolor{bestgreen}\textbf{66.0} & \underline{72.0} & 31.0 & 58.0 & 54.7 & +11.9$^{***}$ \\
\multirow{-4}{*}{\rotatebox[origin=c]{90}{\scriptsize\textit{\begin{tabular}{@{}c@{}}Non-VLM\\lookup\end{tabular}}}} & DINOv2 1-NN & -- & 28.0 & 36.0 & 61.0 & \cellcolor{bestgreen}\textbf{64.0} & 74.0 & 49.0 & 57.0 & 69.0 & 29.0 & 58.0 & 52.5 & +9.7$^{***}$ \\
\cmidrule(l){1-15}
 & SAGE (KB only) & Sonnet & 22.0 & 33.0 & 59.0 & 37.0 & 66.0 & 58.0 & 46.0 & 43.0 & 22.0 & 42.0 & 42.8 & \textit{ref.} \\
 &  & Haiku & 21.0 & 29.0 & 61.0 & 37.0 & 35.0 & 43.0 & 43.0 & 39.0 & 22.0 & 40.0 & 37.0 & $-$5.8$^{***}$ \\
 & Labels only & Sonnet & \underline{38.0} & \cellcolor{bestgreen}\textbf{46.0} & \underline{74.0} & 54.0 & \underline{80.0} & \underline{68.0} & 50.0 & 60.0 & 32.0 & 56.0 & \underline{55.8} & \underline{+13.0$^{***}$} \\
 &  & Haiku & 27.0 & 42.0 & 71.0 & 46.0 & 64.0 & 57.0 & 49.0 & 51.0 & 29.0 & 55.0 & 49.1 & +6.3$^{***}$ \\
 & \cellcolor{oursrow}\textbf{MACE} & \cellcolor{oursrow}Sonnet & \cellcolor{bestgreen}\textbf{45.0} & \cellcolor{oursrow}\underline{45.0} & \cellcolor{bestgreen}\textbf{77.0} & \cellcolor{oursrow}54.0 & \cellcolor{bestgreen}\textbf{84.0} & \cellcolor{bestgreen}\textbf{73.0} & \cellcolor{oursrow}\underline{63.0} & \cellcolor{bestgreen}\textbf{77.0} & \cellcolor{bestgreen}\textbf{34.0} & \cellcolor{oursrow}70.0 & \cellcolor{bestgreen}\textbf{62.2} & \cellcolor{bestgreen}\textbf{+19.4$^{***}$} \\
\multirow{-6}{*}{\rotatebox[origin=c]{90}{\scriptsize\textit{\begin{tabular}{@{}c@{}}Frozen frontier\\VLMs (Claude)\end{tabular}}}} & \cellcolor{oursrow} & \cellcolor{oursrow}Haiku & \cellcolor{oursrow}32.0 & \cellcolor{oursrow}41.0 & \cellcolor{oursrow}71.0 & \cellcolor{oursrow}55.0 & \cellcolor{oursrow}78.0 & \cellcolor{oursrow}63.0 & \cellcolor{oursrow}57.0 & \cellcolor{oursrow}66.0 & \cellcolor{oursrow}\underline{33.0} & \cellcolor{oursrow}59.0 & \cellcolor{oursrow}55.5 & \cellcolor{oursrow}+12.7$^{***}$ \\
\midrule
\multicolumn{15}{l}{\cellcolor{gMean}\textbf{Held-out collection (HO) queries}} \\
 & Label majority vote & -- & 16.7 & 23.3 & 46.7 & 65.0 & 53.3 & 26.7 & 30.0 & 73.3 & 31.7 & 48.3 & 41.5 & $-$9.2$^{**}$ \\
 & Score-weighted vote & -- & 18.3 & 26.7 & 45.0 & \cellcolor{bestgreen}\textbf{70.0} & 58.3 & 33.3 & 31.7 & \underline{78.3} & 36.7 & 46.7 & 44.5 & $-$6.2$^{*}$ \\
 & OpenCLIP 1-NN & -- & 13.3 & 30.0 & 35.0 & 58.3 & 48.3 & 31.7 & 25.0 & 76.7 & 31.7 & 26.7 & 37.7 & $-$13.0$^{***}$ \\
\multirow{-4}{*}{\rotatebox[origin=c]{90}{\scriptsize\textit{\begin{tabular}{@{}c@{}}Non-VLM\\lookup\end{tabular}}}} & DINOv2 1-NN & -- & 13.3 & 26.7 & 40.0 & 58.3 & 45.0 & 33.3 & 31.7 & 71.7 & 35.0 & 26.7 & 38.2 & $-$12.5$^{***}$ \\
\cmidrule(l){1-15}
 & SAGE (KB only) & Sonnet & 15.0 & \cellcolor{bestgreen}\textbf{46.7} & 63.3 & 33.3 & \underline{68.3} & \cellcolor{bestgreen}\textbf{66.7} & 70.0 & 53.3 & 38.3 & 51.7 & 50.7 & \textit{ref.} \\
 &  & Haiku & 23.3 & 40.0 & 60.0 & 43.3 & 21.7 & 51.7 & \cellcolor{bestgreen}\textbf{76.7} & 55.0 & 15.0 & 38.3 & 42.5 & $-$8.2$^{***}$ \\
 & Labels only & Sonnet & 16.7 & 40.0 & \cellcolor{bestgreen}\textbf{68.3} & 65.0 & \cellcolor{bestgreen}\textbf{71.7} & \underline{63.3} & 63.3 & 56.7 & \underline{41.7} & \underline{61.7} & \underline{54.8} & \underline{+4.2$^{*}$} \\
 &  & Haiku & \underline{25.0} & \cellcolor{bestgreen}\textbf{46.7} & \cellcolor{bestgreen}\textbf{68.3} & 48.3 & 38.3 & 50.0 & \underline{73.3} & 66.7 & 26.7 & 60.0 & 50.3 & $-$0.3 \\
 & \cellcolor{oursrow}\textbf{MACE} & \cellcolor{oursrow}Sonnet & \cellcolor{oursrow}20.0 & \cellcolor{bestgreen}\textbf{46.7} & \cellcolor{oursrow}61.7 & \cellcolor{oursrow}\underline{66.7} & \cellcolor{oursrow}61.7 & \cellcolor{oursrow}61.7 & \cellcolor{oursrow}66.7 & \cellcolor{oursrow}\underline{78.3} & \cellcolor{bestgreen}\textbf{43.3} & \cellcolor{bestgreen}\textbf{63.3} & \cellcolor{bestgreen}\textbf{57.0} & \cellcolor{bestgreen}\textbf{+6.3$^{***}$} \\
\multirow{-6}{*}{\rotatebox[origin=c]{90}{\scriptsize\textit{\begin{tabular}{@{}c@{}}Frozen frontier\\VLMs (Claude)\end{tabular}}}} & \cellcolor{oursrow} & \cellcolor{oursrow}Haiku & \cellcolor{bestgreen}\textbf{28.3} & \cellcolor{oursrow}\underline{43.3} & \cellcolor{oursrow}\underline{66.7} & \cellcolor{oursrow}63.3 & \cellcolor{oursrow}31.7 & \cellcolor{oursrow}58.3 & \cellcolor{oursrow}71.7 & \cellcolor{bestgreen}\textbf{81.7} & \cellcolor{oursrow}33.3 & \cellcolor{oursrow}56.7 & \cellcolor{oursrow}53.5 & \cellcolor{oursrow}+2.8 \\
\bottomrule
\end{tabular}}
\end{table*}

\section{Results}

\paragraph{Accuracy across crops.} MACE improves over SAGE with both diagnosis models on both splits, and the gain is largest in distribution (Table~\ref{tab:hero}). Against its own SAGE baseline, Haiku gains 18.5 points IID and 11.0 held-out (both $p{<}0.001$), and every $\Delta$ in Table~\ref{tab:hero} is computed against Sonnet SAGE. In distribution, MACE raises accuracy on all ten crops with both models, with every per-crop 95\% interval above zero (sign test $p{=}0.002$), so the benefit is not limited to one crop group. The largest in-distribution gains appear for mango, cucumber, and corn with Sonnet and for potato and mango with Haiku. MACE also exceeds DINOv2 1-NN with Sonnet on both splits and with Haiku on held-out collections, whereas the Haiku in-distribution comparison remains inconclusive.

In Table~\ref{tab:hero}, the label-lookup baselines exceed SAGE in distribution but fall below it on held-out collections, as expected from their reliance on near neighbors from the same collections. SAGE is more accurate on held-out collections than in distribution, consistent with its KB not depending on the memory images. With Sonnet, MACE exceeds every lookup baseline in distribution and SAGE on held-out collections. Labels only falls between SAGE and MACE on both splits, and Haiku follows the same ordering at lower accuracy.

On held-out collections, the gain depends on the source of the held-out images. For rice, mango, and tomato, whose held-out images do not come from Bugwood, retrieval usually finds the true disease, and MACE gains about 21 points with both models (Sonnet: +21.1 [14.4, 27.8]). On the seven Bugwood crops, Sonnet MACE matches SAGE on average (+0.0 [$-$3.6, +3.6]) and falls below it for wheat, potato, apple, and grape, whereas Haiku gains on these crops on average (+6.4 [1.7, 11.2]) and on every one of them except grape. On Bugwood, a true-disease reference appears among the four for about two-thirds of queries, 18 points [12, 25] fewer than for the other sources. When retrieval misses, MACE falls below SAGE, and with Sonnet, MACE corrects as many SAGE errors on Bugwood as it introduces (31 each). Sonnet SAGE is also more accurate on Bugwood than on the other sources, by 12.9 points [4.4, 21.2], whereas Haiku, whose SAGE baseline is weaker, gains more on the same queries. The low DINOv2 1-NN accuracy on Bugwood suggests that these images differ visually from the memory collections. Because the same crops gain strongly in distribution, the failure is associated with the held-out source rather than with the crop. All seven crops share Bugwood, however, so crop and source effects cannot be separated with these data.

\begin{figure}[h]
\centering
\includegraphics[width=\columnwidth]{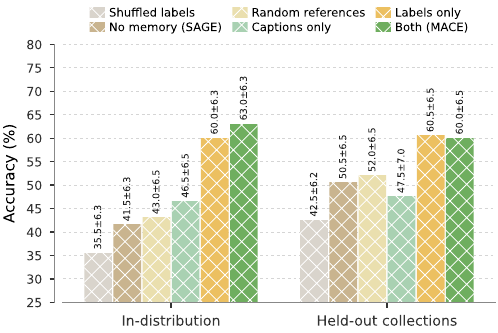}
\caption{\textbf{Sonnet accuracy on 400 ablation queries} by memory content (labels: accuracy $\pm$ half the 95\% interval).}
\label{fig:ablate}
\end{figure}

\paragraph{Labels and captions.} Retrieved labels account for most of the improvement with both models (Table~\ref{tab:hero}; Figure~\ref{fig:ablate}). Captions only and random references remain close to SAGE, while shuffled labels reduce accuracy. On the 400-query ablation set, labels alone match MACE within the 95\% intervals on both splits and equal it on held-out collections. This set is too small to separate captions from labels, whereas the caption effect in distribution is detected on the full query set, analyzed below. The memory therefore helps mainly through its labeled cases, and the descriptions add accuracy mainly when memory and query come from the same collections.

On the full query set, we separate the contribution of captions from the choice of retriever (Table~\ref{tab:factorial}). Caption text adds an IID gain under both BM25 and DINOv2-only retrieval, with no detectable interaction between captions and the retriever. Only the IID caption effects remain significant under the paired bootstrap with Holm correction, and Haiku shows the same IID caption gain under BM25. In contrast, the caption effects on held-out collections remain inconclusive, neither retriever is detectably better, and both effects vary across crops.

\begin{table}[h]
\centering\footnotesize
\caption{\textbf{Sonnet caption and retriever effects} on 1,600 queries (points, 95\% intervals).}
\label{tab:factorial}
\resizebox{\columnwidth}{!}{\begin{tabular}{lcc}
\toprule
Contrast & IID & Held-out \\
\midrule
Captions, BM25 & +6.4 {\scriptsize[+4.1, +8.9]} & +2.2 {\scriptsize[$-$0.8, +5.2]} \\
Captions, DINOv2-only & +6.3 {\scriptsize[+3.8, +8.7]} & +0.5 {\scriptsize[$-$2.7, +3.5]} \\
DINOv2-only $-$ BM25, labels & $-$0.6 {\scriptsize[$-$2.9, +1.8]} & +1.0 {\scriptsize[$-$2.0, +4.0]} \\
DINOv2-only $-$ BM25, captions & $-$0.7 {\scriptsize[$-$3.1, +1.8]} & $-$0.7 {\scriptsize[$-$3.5, +2.2]} \\
Interaction & $-$0.1 {\scriptsize[$-$3.0, +2.7]} & $-$1.7 {\scriptsize[$-$5.7, +2.0]} \\
\bottomrule
\end{tabular}}
\end{table}

\begin{figure}[t]
\centering
\includegraphics[width=\columnwidth]{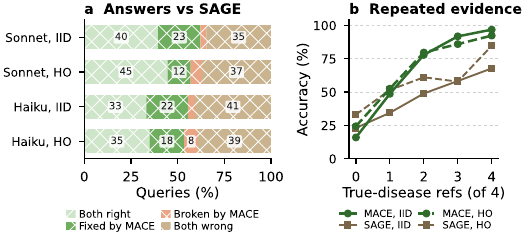}
\caption{\textbf{MACE versus SAGE.} (a) Answers fixed or broken. (b) Sonnet accuracy by true-disease references.}
\label{fig:outcomes}
\end{figure}

\paragraph{Effect on individual predictions.} MACE corrects more SAGE errors than it introduces for both models and on both splits (Figure~\ref{fig:outcomes}a; exact McNemar $p{<}0.001$ in all four model--split pairs). Accuracy rises with the number of retrieved references labeled with the true disease (Figure~\ref{fig:outcomes}b), and SAGE accuracy on the same queries also rises with this number, so part of the trend reflects query difficulty. Even so, the gain over SAGE is 10 to 20 points larger with two or more true-disease references than with one, significantly in three of four model--split pairs under the paired bootstrap. When retrieval omits the true disease, MACE falls below SAGE in all four pairs, significantly in three, and the exception is Sonnet on held-out collections ($-$6.7 [$-$13.6, +0.2]). These findings indicate that the diagnosis model benefits both from receiving the true disease and from receiving it more than once. The same findings explain the Bugwood pattern in Table~\ref{tab:hero}, where frequent misses offset the gains on the remaining queries.

\begin{figure}[h]
\centering
\includegraphics[width=\columnwidth]{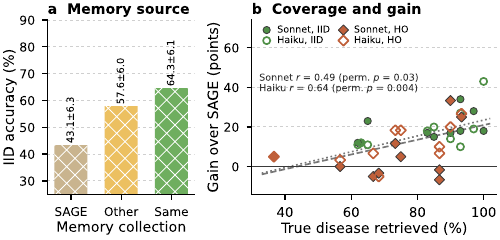}
\caption{\textbf{Memory.} (a) Sonnet IID accuracy by memory source ($\pm$ half the 95\% interval). (b) Gain versus coverage.}
\label{fig:memory}
\end{figure}

\paragraph{How the memory is built.} Memory from the query's own collection gives the highest IID accuracy (Figure~\ref{fig:memory}a), mainly because removing that collection leaves many queries without a reference labeled with their true disease. A smaller advantage remains even when the disease is retained, suggesting that same-collection similarity also helps. Across crops and splits, larger gains over SAGE are associated with higher retrieval coverage for both models (Figure~\ref{fig:memory}b). However, coverage does not explain every result: Sonnet loses accuracy on held-out potato images despite high coverage, while Haiku improves on the same images.

\begin{figure}[h]
\centering
\includegraphics[width=\columnwidth]{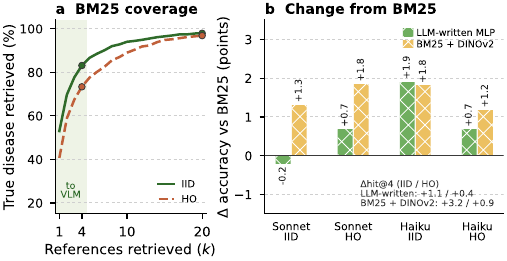}
\caption{\textbf{Selection rules.} (a) BM25 coverage by candidates retrieved. (b) Accuracy change from BM25.}
\label{fig:selection}
\end{figure}

\paragraph{Exploratory evaluation: choosing the four references.} We next ask whether a better rule for choosing the four references can raise accuracy. BM25 places a true-disease reference among its top 20 candidates for nearly every query (Figure~\ref{fig:selection}a), so a better rule can change only which of these candidates enter the final four.

\paragraph{An LLM-written retriever.} Sonnet can write the weights of a retriever without gradient training: in a single call, it writes the 29 parameters of the MLP. All twelve written versions, including the three given only the feature description, place a true-disease reference among the four at least as often as BM25 in distribution, although no version is significantly better under the paired bootstrap. The locked version, chosen on development data, received the description, memory captions, and 16 labeled triplets. The same MLP fitted by gradient descent on those 16 triplets falls about twenty points below BM25 in IID coverage, so writing the weights is far more data-efficient than fitting them on the same examples. In diagnosis, the locked LLM-written retriever stays level with BM25 for Sonnet and gives Haiku an in-distribution gain of about two points that is not significant under the paired bootstrap (Figure~\ref{fig:selection}b). An LLM can therefore write competitive retriever weights without gradient training, although the features, architecture, and prompt were chosen on development data.

\paragraph{Adding image similarity.} Adding DINOv2 image similarity to a retuned BM25 score improves coverage modestly, mainly on in-distribution queries, and the accompanying diagnosis changes are small and positive for both models but not significant after Holm correction (Figure~\ref{fig:selection}b). Both rules mostly reorder references that BM25 already finds. A rule that raises coverage can also lower accuracy. In a targeted check with Sonnet, keeping at most one reference per disease places the true disease among the four more often but lowers IID accuracy by about eight points, with a paired interval that excludes zero. The restriction changes both which labels appear and how often they appear, and it removes the repeated true-disease references that are associated with the steepest accuracy gains in Figure~\ref{fig:outcomes}b. Losing repeated evidence for the correct disease is therefore a plausible explanation, although the change in the selected set may also contribute. The check also covers only one rule on a subset of queries. These findings indicate that hit@4 alone does not determine reference quality and that a selection rule should also be assessed on diagnosis accuracy.

\section{Conclusion}

We introduced MACE, a training-free method that adds a caption memory of labeled reference images to frozen VLM diagnosis. Across ten crops and two frozen frontier models, MACE improves over the source-grounded SAGE baseline on every crop in distribution and on held-out collections on average, with strong variation across held-out sources. With Sonnet, MACE also exceeds visual nearest-neighbor lookup on both splits.

The memory helps mainly as labeled case evidence. Retrieved labels account for most of the gain, caption text adds a further in-distribution improvement, and accuracy rises with the number of retrieved references labeled with the true disease. When retrieval misses the true disease, as it often does for Bugwood images, the memory lowers accuracy, so drawing memory cases from more varied sources is a promising way to improve transfer to new collections.

In an exploratory evaluation, a retriever whose weights Sonnet writes without gradient training matches BM25 in coverage and far exceeds the same model fitted on the same few examples. Image similarity adds only modest coverage, and selection rules should be judged on diagnosis accuracy, not on coverage alone.

\paragraph{Limitations.} Our evaluation covers ten crops, four held-out sources, and two Claude models evaluated once per query, and no plant pathologist reviewed the labels.

\paragraph{Image credits.} Images in Figures~\ref{fig:overview} and~\ref{fig:cropset} were cropped and resized from PlantDoc, LeafNet \citep{quoc2026leafnet}, PlantVillage, MangoLeafBD \citep{ahmed2023mangoleafbd}, and the datasets of \citet{menco2025cucumber} and \citet{raghavendra2024vqa}, together with the Iowa State plant-stress phenotyping collection. Each image remains subject to its source license; the website provides the original source and license for every image.

\section*{Acknowledgments}
This work was supported by COALESCE: COntext Aware LEarning for Sustainable CybEr-Agricultural Systems (NSF \#1954556) and the AI Institute for Resilient Agriculture (USDA-NIFA \#2021-67021-35329).

\bibliography{references}

\end{document}